\documentclass[letter]{aa} 

\usepackage{xcolor}
\usepackage{graphicx}
\usepackage{txfonts}
\usepackage{tikz,lipsum,lmodern}
\usepackage[most]{tcolorbox}
\usepackage{subcaption}
\usepackage{booktabs}
\usepackage[colorlinks=true,linkcolor=purple,citecolor=blue,filecolor=black,urlcolor=cyan,final=true]{hyperref}
\begin{document}

   \title{Relativistic electron beams accelerated by an interplanetary shock}

   \author{Immanuel C. Jebaraj
          \inst{1}
          \and
          N. Dresing\inst{1}
          \and
          V. Krasnoselskikh \inst{2, 3}
          \and
          O. V. Agapitov \inst{3}
          \and
          J. Gieseler \inst{1}
          \and
          D. Trotta \inst{4}
          \and
          N. Wijsen \inst{5}
          \and
          A. Larosa \inst{6}
          \and
          A. Kouloumvakos \inst{7}
          \and
          C. Palmroos \inst{1}
          \and 
          A. Dimmock \inst{8}
          \and
          A. Kolhoff \inst{9}
          \and
          P. K\"uhl \inst{9}
          \and
          S. Fleth \inst{9}
          \and
          A. Fedeli \inst{1}
          \and
          S. Valkila \inst{1}
          \and
          D. Lario \inst{10}
          \and
          Yu. V. Khotyaintsev \inst{8}
          \and
          R. Vainio \inst{1}
          }

   \institute{Department of Physics and Astronomy, University of Turku, FI-20500 Turku, Finland\\
              \email{immanuel.c.jebaraj@gmail.com}
         \and
             LPC2E/CNRS, UMR 7328, 3A Avenue de la Recherche Scientifique, Orleans, France
         \and
             Space Sciences Laboratory, University of California, Berkeley, CA, USA
         \and
             The Blackett Laboratory, Department of Physics, Imperial College London, London, UK
         \and
             Center for mathematical Plasma Astrophysics, Department of Mathematics, KU Leuven, Celestijnenlaan 200B, B-3001 Leuven, Belgium
         \and
             Queen Mary University of London, School of Physics and Astronomy, London, UK
         \and
             The Johns Hopkins University Applied Physics Laboratory, Laurel, MD 20723, USA
         \and
             Swedish Institute of Space Physics, P.O. Box 537, SE-751 21 Uppsala, Sweden
         \and
             Institute of Experimental \& applied Physics, Kiel University, 24118 Kiel, Germany
         \and
             Heliophysics Science Division, NASA Goddard Space Flight Center, Greenbelt, MD 20771, USA
             }

   \date{}

 
  \abstract
   {Collisionless shock waves have long been considered amongst the most prolific particle accelerators in the universe. Shocks alter the plasma they propagate through and often exhibit complex evolution across multiple scales. Interplanetary (IP) traveling shocks have been recorded in-situ for over half a century and act as a natural laboratory for experimentally verifying various aspects of large-scale collisionless shocks. A fundamentally interesting problem in both helio and astrophysics is the acceleration of electrons to relativistic energies ($> 300$ keV) by traveling shocks.}
   {The reason for an incomplete understanding of electron acceleration at IP shocks is due to scale-related challenges and a lack of instrumental capabilities. This letter presents first observations of field-aligned beams of relativistic electrons upstream of an IP shock observed thanks to the instrumental capabilities of Solar Orbiter. This study aims to present the characteristics of the electron beams close to the source and contribute towards understanding their acceleration mechanism.}
   {On 25 July 2022, Solar Orbiter encountered an IP shock at 0.98 AU. The shock was associated with an energetic storm particle event which also featured upstream field-aligned relativistic electron beams observed 14 minutes prior to the actual shock crossing. The distance of the beam's origin was investigated using a velocity dispersion analysis (VDA). Peak-intensity energy spectra were anaylzed and compared with those obtained from a semi-analytical fast-Fermi acceleration model.}
   {By leveraging Solar Orbiter's high-time resolution Energetic Particle Detector (EPD), we have successfully showcased an IP shock's ability to accelerate relativistic electron beams. Our proposed acceleration mechanism offers an explanation for the observed electron beam and its characteristics, while we also explore the potential contributions of more complex mechanisms.}
   {}

   \maketitle
%

\section{Introduction}

Shock waves are ubiquitous in space plasmas and are the most prolific particle accelerators in most systems. They can be directly probed in the heliosphere due to the presence of planetary bow shocks (created due to the planetary obstacles in the solar wind flow), and interplanetary shocks which are driven by solar activity such as coronal mass ejections (CMEs). CMEs are large-scale eruptions of plasma and magnetic fields which travel away from the Sun that when their speed exceeds the information speed of the medium (i.e. the fast-magnetosonic speed), shock waves are generated. While ion acceleration in such shocks has been extensively studied \citep[][and references therein]{Lee12}, electron acceleration is challenging due to the large scale separation\footnote{Ratio of gyro-radii ($r_L$) $\sim \sqrt{\frac{m_i}{m_e}} = 42.8$, where $m_i$ and $m_e$ are the mass of protons and electrons, respectively.} and physical limitations such as their retention at the shock and magnetization \citep[see,][for review]{Lembege04}. However, observational surveys have found a small number of interplanetary (IP) shocks that were associated with a significant increase in electron fluxes from sub-relativistic to relativistic energies \citep[e.g.,][]{Sarris85, Lopate89, Dresing16, Grant21, Nasrin23}. Furthermore, a strong correlation was found between the characteristics of relativistic ions and electrons, suggesting potential similarities in their acceleration mechanisms \citep[][]{Posner07, Dresing22}. While acceleration of sub-relativistic electrons ($< 10$ keV) by IP shocks is widely acknowledged due to the common occurrence of type II radio emissions \citep[][]{Krasnoselskikh85a, Bale99, Magdalenic14, Jebaraj20, Kouloumvakos21, Jebaraj21}, acceleration of relativistic electrons ($>300$ keV) by IP shocks remains an open question.




The characteristics and dynamics of collisionless shock waves are complex and are defined across multiple scales \citep[][]{Galeev63,Karpman64, Sagdeev66, Kennel85, Krasnoselskikh13}. These complexities are further enhanced when addressing the behavior of electrons at shock fronts \citep[][]{Balikhin89, Trotta19,Agapitov23}. An important property of shocks is their geometry, that is the angle between the upstream magnetic field ($\textbf{B}$) and the shock normal ($\hat{\textbf{n}}$), $\theta_{Bn}$. Quasi-perpendicular shocks ($\theta_{Bn}>45^{\circ}$) are often considered more efficient in accelerating electrons to high energies \citep[][]{Schatzman63,Bulanov90, Mann09}. The typical width of these shocks are larger than the electron gyro-radius \citep[][]{Walker99, Hobara10} and therefore their rapid acceleration can be explained via two main mechanisms. The first is the ``fast-Fermi'' process, where acceleration happens via magnetic mirroring \citep[due to the steep magnetic gradient at quasi-perpendicular shocks,][]{Leroy84, Wu84}. This mechanism has been exploited for almost four decades when discussing the so-called ``field-aligned beams'' (FABs hereon) of energetic particles \citep[e.g.,][]{Pulupa08}. The second process is a gradient drift mechanism commonly known as ``shock-drift acceleration'' \citep[SDA;][]{Hudson65, Ball2001} which is similar to fast-Fermi acceleration, but deviates due to the fact that the electron drift along the small-scale electric field gradients within the shock ramp \citep[][]{Krasnoselskikh02, Vasko18, Dimmock19, Hanson20}. 


Spacecraft upstream of the terrestrial bow shock have routinely measured $10 - 100$ keV (near-relativistic electrons) \citep[e.g.,][]{Lynn16} and far higher energy electrons at other planetary bow shocks such as those of Jupiter and Saturn \citep[e.g.,][]{Masters17}. The most frequently observed similarity in conditions across all these planetary bow shock observations is the presence of a shock that has exceeded a certain critical point beyond which the downstream sound speed ($c_s$) is greater than the flow speed. Such shocks are termed, ``super-critical'' and behave fundamentally differently from those where $c_s$ is slower than the flow speed downstream (sub-critical). This is characterized best by the ratio between the shock speed and the fast-magnetosonic speed known as the Mach number $M$. Super-critical shocks are usually strong shocks, while sub-critical are weak shocks. Depending on the shock geometry, super-critical shocks can generate ion-kinetic structures in the region ahead of the shock (foreshock), which trap and energize particles \citep[][]{Kennel81, Kis13}. However, observations of electron trapping and energization are rare in IP shock waves, which are generally weak shocks \citep[][]{Armstrong76, Sarris85, Shimada99}. Super-critical shocks may also be crucial for ion acceleration and may be the reason for the good correlations between relativistic ions and electrons at strong coronal shocks \citep[][]{Kouloumvakos19, Dresing22}.

In this study, we present measurements of relativistic electron beams found upstream of a travelling shock wave driven by an IP coronal mass ejection (ICME). To our knowledge, this is the first such study made possible thanks to the enhanced time resolution (1 second) of the Energetic Particle Detector \citep[EPD;][]{Pacheco2020} instrument suite onboard Solar Orbiter \citep[SolO; ][]{Muller13}. We utilized the Electron Proton Telescope (EPT) which covers electron energy ranges from $\approx 30$ keV -- $300$ keV, and the High Energy Telescope (HET) which covers $> 400$ keV. The letter is structured as follows; An overview of the event is presented in Sect. \ref{Sec:observations}, experimental details of the FABs and a demonstration of their origin from a remote location of the shock are presented in Sect.~\ref{Sec:electron_obs}, and a robust peak-intensity spectral analysis in Sect.~\ref{Sec:energy_spectra}. Finally, a simple shock acceleration model for FABs is presented in Sect.~\ref{Sec:shock_model}, before discussing the results and its implications in Sect.~\ref{Sec:discussions}

\section{Overview of the insitu shock and associated energetic electrons} \label{Sec:observations}

 \begin{figure*}
 \centering
    \includegraphics[width=0.91\textwidth]{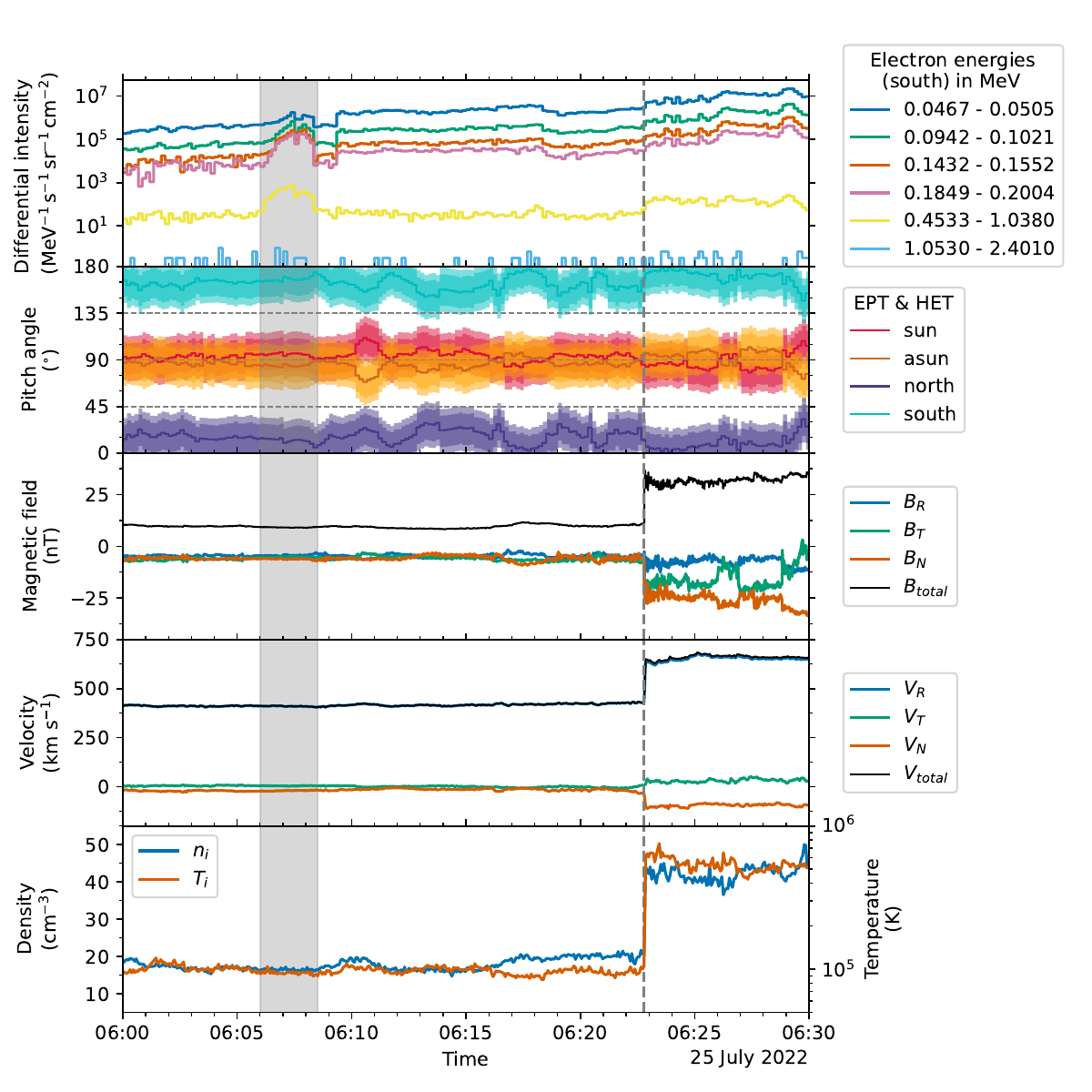}
  \caption{Overview of the in-situ shock and energetic electrons observed by Solar Orbiter on 25 July 2022. The arrival of the shock at 06:22:47 is marked by the vertical, gray dashed line in all panels.
  The \textit{first} panel shows the differential intensities of energetic electrons observed by the south telescopes of EPT and HET (covering a stable pitch angle of $~180^{\circ}$ throughout the shown period). 
  The upstream FABs are observed between 06:06:00--06:08:30~UT and indicated by the gray shaded region. Fig.~\ref{Fig_app:full_electrons} shows the differential intensities across all viewing directions.
  The \textit{second} panel presents the pitch-angle coverages for the different viewing directions of EPT and HET, where the gradients in color indicate the different telescope openings of EPD/EPT (smaller) and EPD/HET (larger). 
  The \textit{third} \& \textit{fourth} panels show the components of the magnetic field, and bulk flow velocity, while the \textit{fifth} panel presents the ion number density and proton temperature. }
  \label{Fig:overview}
  \end{figure*}

On 25th July 2022, at 06:22:45 UT, the SolO spacecraft encountered an IP fast-forward shock wave as seen in the last three panels of Fig.~\ref{Fig:overview}. It was linked to a weak C1 solar flare and CME on 23rd July 2022 at about 18:30~UT. A remarkable aspect related with the in-situ shock crossing, as shown in the first two panels of Fig.~\ref{Fig:overview} (and Fig.~\ref{Fig_app:full_electrons}), is the presence of upstream field-aligned beams of relativistic electrons (up to $\sim 1$~MeV) marked by the gray shaded region. The shock was also associated with an energetic storm particle (ESP) event for electrons, the features of which are outside the scope of this letter, but a short summary is provided in Appendix~\ref{app:all_telescopes}.
 

The shock arrived at SolO approximately 36 hours after the solar event, which results in a average transit speed of 1100 km s$^{-1}$. In order to obtain the in-situ characteristics of the shock wave, such as its speed and strength, a more robust physical approach involving multiple steps was adopted.

Firstly, the shock normal ($\hat{\textbf{n}}$) was estimated using a combination of the mixed-mode method, magnetic and velocity coplanarity \citep[][and references therein]{Trotta22} and minimum variance analysis \citep[MVA; ][]{Sonnerup98} to be $\hat{\textbf{n}} = [0.93 \pm 0.05, 0.018 \pm 0.12, -0.16 \pm 0.25]$. Next, the shock's obliquity was estimated to be $\theta_{Bn} \sim 64^{\circ} \pm 6^{\circ}$ indicating a quasi-perpendicular geometry. The upstream and downstream bulk flow speeds were then estimated in the shock rest frame, $V_{u}^{sh} \sim 350 \pm 50$ km s$^{-1}$ and $V_{d}^{sh} \sim 125 \pm 50$ km s$^{-1}$, which results in a shock speed, $V_{sh} \sim 775 \pm 50$ km s$^{-1}$ in the observers frame. The Alfvén Mach number ($M_A$) was then estimated using the upstream Alfvén speed, $c_A = B_{up}/\sqrt{\mu_0 n_i m_i}$ (assuming $Z=1$, $m_i$ and $n_i$ are proton mass and number density), and proxies (for $M_A$) established in \cite{Gedalin21} to be $M_A \sim 3.3 \pm 0.7$. Single spacecraft techniques to obtain shock parameters are cumbersome and may possess large uncertainties and proxies act as constraints. Another indicator of the shock's strength is its compression ratios: $r_{B}\sim$3 for magnetic compression ($B_{down}/B_{up}$) and $r_{gas}$ of at least $\sim$2.1 for gas compression ($n_{down}/n_{up}$). The estimated $r_{gas}$ may be influenced by data uncertainties and therefore for simplicity, $r_{gas}$ is ``at least'' $\sim 2.1$ as for a quasi-perpendicular shock, $r_{gas}$ is expected to be similar to $r_B$. This uncertainty was also considered when estimating $V_{sh}$ and $M_A$. 

\section{Upstream field-aligned electron beams} 

\subsection{Observational details and the properties of the electron beams} \label{Sec:electron_obs}

During the period of shock arrival (c.f. Fig. \ref{Fig:overview}), the pitch-angle (PA) coverage of EPD was unusually stable due to the very stable magnetic field vector. The orientation of the magnetic field was north-south, resulting in field-aligned anti-sunward propagating particles being detected in the south telescopes of SolO/EPD instruments with PA close to 180$^{\circ}$. Consequently, the north telescope measured particles propagating toward the Sun, while the sun and anti-sun telescopes covered particle PA perpendicular to the magnetic field ($\sim90^\circ$). Taking this into account, electron observations from only the south telescope are presented in the main text. Detailed observations from all telescopes can be found in Appendix \ref{app:all_telescopes}. Since the measured electron intensities may be subject to ion contamination, robust contamination corrections were applied using different techniques for both EPT (Appendix \ref{app:conta_corr}) and HET (Appendix \ref{app:het}) measurements. 

The first panel in Fig.~\ref{Fig:overview} presents the differential electron intensities observed by the south telescopes of EPT and HET post-correction, which show a clear increase in the electrons above the background. Thanks to the high time resolution provided by EPT and HET (full resolution is 1 second), it was possible to identify, for the first time, fine features, particularly, electron FABs upstream of the shock between 06:06:00 and 06:08:30~UT ($\sim$ 14 minutes prior to shock arrival) in the south facing telescopes (indicated by the gray shaded region in Fig.~\ref{Fig:overview}). Such FABs have never been clearly measured upstream of interplanetary shocks before, possibly due to the lack of sufficient time resolution of the particle measurements. On closer inspection, this particular FAB had two peaks within it which are distinguishable over the same range of energies. Since the FABs were only observed by the south telescopes, and the lower-energy STEP telescope only has a field of view in the sunward direction, the FABs could not be observed in lower energies (sub-relativistic, i.e. $<10$ keV). However, Langmuir wave packets were registered by the Time Domain Sampler (TDS) instrument, which is a part of the Radio Plasma Waves instrument suite \citep[RPW;][]{Maksimovic20b}. Two wave packets (not shown), first at 06:10:05 UT and then 06:11:59 UT, are likely evidences of the FABs extending to sub-relativistic energies. Other than small amplitude Langmuir waves, no strong wave activity was measured during this time, making it improbable for the FABs to be generated locally. The presence of Langmuir waves also indicates the remote origin of the beam, as it must have propagated a distance, at the very least, of the order of its relaxation length \citep[][]{Ryutov69, Vosh15, Jebaraj23}.

Under the assumption that electrons of all energies were injected at the same time \citep[e.g.,][]{Vainio13}, a velocity dispersion of the FABs allows to estimate the injection time and the propagated path length of the FABs. The details of this analysis can be found in Appendix \ref{app:vda}. The results of the analysis shown in Fig. \ref{Fig:VDA_temp} suggest that the electrons were accelerated in a region $\sim$ 13 $R_{\odot}$ away from the observer and were injected at 06:05:46 UT.

\subsection{Assessing the characteristics of electron acceleration using peak energy spectra} \label{Sec:energy_spectra}

 \begin{figure*}
 \centering
    \includegraphics[width=0.95\textwidth]{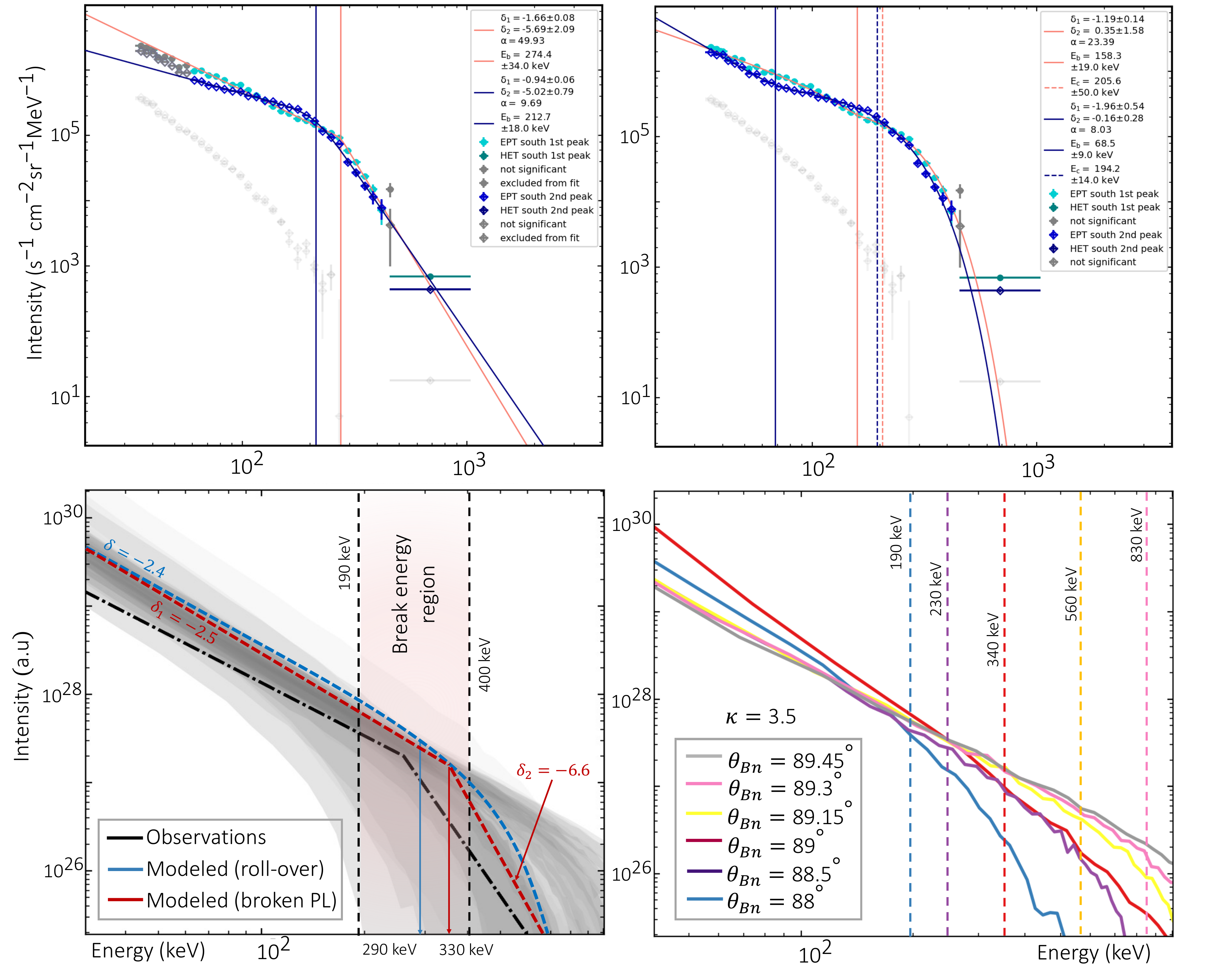}
  \caption{Observation (\textit{top}) vs model (\textit{bottom}) comparison of the peak-intensity electron spectra. In the upper two panels, points in aqua signify the first peak, while points in blue denote the second peak. The corresponding fits are indicated by orange for the first peak and blue for the second. The background that has been subtracted from each energy channel is represented by points in gray.(\textit{top-left}) Broken power-law fits applied to both peaks. (\textit{top-right}) Broken power-law fits with an exponential roll-over (or cutoff) for the two peaks. (\textit{bottom-left}) Ensemble of runs indicated by the gray shaded region with varying $\kappa \sim 3 - 4$ and $\theta_{Bn} \sim 87^{\circ} - 89.5^{\circ}$. A single power-law fit and exponential roll-over (blue dashed line), and a broken power-law fit (red dashed line) are shown for the modeled mean. The fit for the first peak from top-left panel is illustrated by the olive-black dash-dot line. The mean roll-over ($E_c$) and break energies ($E_b$) of the ensemble and its standard deviation are also shown. (\textit{bottom-right}) Modeled results using $\kappa \sim 3.5$ for different $\theta_{Bn}$ ($88^{\circ} - 89.5^{\circ}$). The corresponding break energies are marked by the vertical dashed lines.}
  \label{Fig:obs_model}
  \end{figure*}

The generation and propagation of the FABs can be understood better by constructing a peak-intensity spectrum across all observed energy channels \citep[e.g.,][]{Dresing2020}. The two top panels of Fig. \ref{Fig:obs_model} present background-subtracted and contamination corrected electron peak-intensity spectra of the two well-defined peaks within the FAB, using measurements from both EPT and HET south-facing telescopes and a time resolution of 15 seconds. Light gray markers represent the subtracted background, while the colored markers indicate the peak electron fluxes. 


The method for spectral fitting outlined in \cite{Dresing2020, Strauss2020} allows for different fit models. For the FABs, we use a broken power law defined by two different spectral indices $\delta_1$ and $\delta_2$ separated at a break-energy $E_b$, a single power law ($\delta_1$) with an exponential roll-over or cutoff energy $E_c$, and a broken power law with an exponential roll-over at higher energies. Depending on the respective peak and the energy range taken into account in the fitting procedure, different models that fit the data were obtained. The top-right panel of Fig.~\ref{Fig:obs_model} shows the fit applied to both peaks over all available EPT channels and therefore fitting also the ``ankle-like'' feature below $100$~keV, which is more pronounced in the spectrum of the second peak than the first. The resulting fits both yield a similar exponential roll-over at higher energies. If the ankle is excluded from the fit as done  in Fig. \ref{Fig:obs_model} (top-left), the fit results in a broken power law for both peaks. Although these fits seem to yield a better agreement with HET points, it is hard to conclude what type of fit, namely, a broken power law or an exponential roll-over is the best representation of the spectrum. This is on the one hand due to the HET electron peak fluxes potentially being subject to a small amount of proton contamination, which would lead to slightly increased fluxes, and on the other hand because of the ion-contamination correction, which was applied to the EPT data points, not being perfect. The presence of a break or spectral roll-over is without doubt, as it is already evident in the uncorrected EPT data (not shown).

The cutoff energies, $E_c$ for both the fits on the right panel are similar, $\sim 200$ keV. However, the break energies, $E_b$ for the fits in the left panel are both found above $\sim200$ keV, but are slightly different. The most significant difference in the two spectra was at lower energies, the second peak exhibits a more pronounced ankle at approximately $\sim$ 70 keV. The lower spectral index of the first peak in the right panel is $\delta = -1.19\pm0.14$, while that of the second peak is $\delta = -1.96\pm0.54$. For comparison, the spectral indices for the first peak in the left panel are, $\delta_1 = -1.66\pm0.08$ and $\delta_2 = -5.69\pm2.09$, while that of the second peak are $\delta_1 = -0.94\pm0.06$ and $\delta_2 = -5.02\pm0.79$, respectively. We also fit an averaged spectrum of a time period of 2.5 min containing both peaks, which results in a broken power law with $\delta_1 = -1.4\pm0.04$, $\delta_2 = -5.2\pm2.58$ and $E_b = 261.8\pm47$ keV (not shown).



\section{A model for electron acceleration at IP shocks} \label{Sec:shock_model}

From the observations presented in this letter, it is understood that the FABs were accelerated at a remote location on the shock similar to how they are accelerated at the near-perpendicular terrestrial bow shock. If so, then the curvature of the shock (or large scale deformations for IP shock at 1 AU) and the time-varying properties of the upstream magnetic field become crucial. Acceleration happens under the exclusive assumption that there exists a region on the shock wave separated by $\sim 13 R_{\odot}$ that is near-perpendicular ($89.9^{\circ}>\theta_{Bn}>85^{\circ}$) and super-critical \footnote{When the downstream medium is subsonic, i.e. the sound speed $c_s > V_d^{sh}$, the shock undergoes fundamental changes in its evolution and structure \citep[][]{Krasnoselskikh85b}} (Appendix \ref{app:fast_fermi}). When these conditions are met, acceleration happens based on the conservation of magnetic moment \citep[fast-Fermi mechanism,][]{Sonnerup69, Wu84, Leroy84} through which electrons can gain a significant amount of energy in a single encounter with a near-perpendicular shock.  


A primary aspect of super-critical shocks is that in order to dissipate excess energy, they reflect a portion of upstream ions forming an overshoot magnetic field \citep[][]{Krasnoselskikh13}. The steep magnetic gradient of the overshoot enables the near-relativistic and relativistic electrons to conserve their magnetic moment and as such, a relativistic approach to the fast-Fermi mechanism must be adopted (Appendix \ref{app:fast_fermi}). It is noteworthy to mention that a sub-critical shock is also capable of reflecting electrons but to a far lesser degree as far less particles would participate in such a process. The second important assumption is that in a near-perpendicular geometry, the shock speed along the magnetic field line scales as a function of geometry. This limits acceleration to only those particles that exceed it, resulting in dilute beams.

We perform semi-analytical modeling based on the relativistic form of the reflection conditions provided in \cite{Leroy84}. When \(\theta_{Bn} > 89.5^{\circ}\), the shock speed must also be Lorentz transformed as it is relativistic. The upstream ``seed'' electron distribution function was assumed to be a kappa-distribution (Appendix \ref{app:fast_fermi}) with $\kappa \sim 3 - 4$ \citep[considered realistic, ][]{KraussV1991}. For the parameters of the shock, an upstream magnetic field similar to the one measured in-situ by SolO was used, i.e. $\sim$10 nT. The magnetic compression including the overshoot magnetic field is considered to be $B_{overshoot}/B = 7$ \citep[based on, ][]{Mellott87} yielding a maximum $B$ of 70 nT. Lastly, the effect of the electrostatic potential cannot be ignored for electrons and as such, 100 eV was used (obtained through proxies, Appendix \ref{app:fast_fermi}).

The modeling results are presented in the bottom two panels of Fig. \ref{Fig:obs_model}. An ensemble of runs was performed using kappa values between $\kappa \sim 3 - 4$, and shock obliquity between $\theta_{Bn} \sim 87^{\circ} - 89.5^{\circ}$. The gray shaded region in the bottom-left panel of Fig.\ref{Fig:obs_model} shows the results and two different fits to the mean of all runs, namely, a single power-law with a roll-over (blue dashed line), and a broken power-law (red dashed line). The first has a power-law with spectral index $\delta \sim - 2.4$ which is followed by an exponential roll-over, while the second has two power-laws with spectral indices $\delta_1 \sim - 2.5$ and $\delta_2 \sim - 6.6$. Both fits are used, but the $\chi^2$ of the first fit is lower than that of the second. The mean cutoff energy ($E_c$) for the first fit was $\sim 290$ keV and the mean break-energy ($E_b$) of the second fit was $\sim 330$ keV. The vertical red shaded region shows the $2\sigma$ standard deviation of both fits which are similar. For comparison, the broken power-law of the first peak (i.e., $\delta_1 \sim -1.66$, and $\delta_2 \sim 5.7$, marked by the orange fit in top-left panel of Fig. \ref{Fig:obs_model}) from observations is represented in black. Both the modeled spectral indices and the break-energy range are comparable to the observations. 


In a general sense, fast-Fermi acceleration simply shifts the seed distribution minus the transmitted electrons to higher energies (see, Appendix \ref{app:fast_fermi}). This is also corroborated by the fact that a single power-law with an exponential roll-over produces a better $\chi^2$ than a broken power-law fit. The mechanism does not change the shape of the source $\kappa$-distribution function. The shift (i.e. energy gain) itself is highly dependent on the shock obliquity, and this is clearly shown in the bottom-right panel of Fig.\ref{Fig:obs_model}. When $\kappa = 3.5$ is a fixed value and the obliquity is varied between $\theta_{Bn} = 88^{\circ} - 89.5^{\circ}$, the resulting energy spectra are significantly different for energies above 100 keV. The difference is quantified by the dotted vertical lines representing the break energies ($E_b$) for each $\theta_{Bn}$. The effect of a changing $\kappa$ index would then be responsible for the steepness of the spectrum since it determines the number of electrons in the tail. A $\kappa = 2$ can be characterized with a power-law index of $\delta \sim -1.5$ at suprathermal energies and an exponential roll-over above a certain energy range \citep[][]{Oka18}. Then, the steepest result possible is when the distribution tends towards a Maxwellian distribution as $\kappa \rightarrow \infty$. 

Finally, it is worth mentioning that the model used here makes two main simplifications, namely, a 1D shock profile, and that the electron's magnetic moment is conserved. In reality, super-critical shocks are complex \citep[][]{Krasnoselskikh85b, Lembege92, Balikhin97}, and acceleration by multiple reflections and subsequent gradient drift (SDA) may also take place (see, Appendix \ref{app:fast_fermi} for more details). In such cases the resultant distribution will deviate from what is presented here and may resemble a broken power-law.


\section{Discussion of the results} \label{Sec:discussions}

On 25 July 2022 a travelling IP shock wave accompanied by an energetic storm particle event was recorded in-situ by Solar Orbiter. Notably, field-aligned beams of relativistic electrons were observed 14 minutes prior to shock arrival. These are the first such observations of electron FABs measured at IP shocks thanks to the high time-resolution of the EPD instrument suite onboard SolO. This letter presents their observational characteristics and proposes a scenario for their acceleration at IP shocks.

A combination of a lack of upstream magnetic structures, and the velocity dispersion of the FABs lead to the conclusion that they originated at a remote location of the shock. Velocity dispersion analysis suggested that the beams originated $\sim$13$ R_{\odot}$ away from the spacecraft. Further corroboration was provided by the presence of plasma waves \citep[Langmuir waves, ][]{Filbert79, Kasaba00}, which are a consequence of beam propagation and relaxation \citep[quasi-linear relaxation,][]{Vedenov62b}.


\cite{Galeev63} initially postulated that shocks evolving in a low-density, inhomogeneous plasma might develop decay instabilities that deform the laminar shock front. In higher dimensions, large-scale deformations lead to curved fronts on the scale of the upstream inhomogeneities, spanning several \( R_{\odot} \) to tens of \( R_{\odot} \) \citep[e.g.,][]{Wijsen23b} which play a significant role in FAB acceleration \citep[][]{Bulanov86, Decker90, Bulanov99, Lario08}. With any curved front, a finite region exists where \( \theta_{Bn} \geq 85^{\circ} \) and approaches \( 90^{\circ} \), allowing for efficient electron acceleration along the tangent magnetic field line as demonstrated in this letter. When the observer is connected to the tangent field line, the time-varying properties of the FABs are likely influenced by both deformation induced curvature and variations in the upstream field topology \citep[e.g.][]{Giacalone17, Trotta23MNRAS}.


Similarities between the observed and modeled peak-intensity spectra of the FABs serve as a demonstration of shock acceleration at an IP shock under the right circumstances. The model naturally forms a roll-over or break at higher energies ($> 200$ keV), which is also seen in observations. If a broken power-law were fit to the modeled results, the break-energy, $E_b$ would fall between 190 keV and 400 keV comparable to the observations. This is considerably different from the transport-related breaks discussed in previous studies \citep[][]{Dresing21}, which have suggested that the presence of two different spectral breaks in peak-intensity spectra of solar energetic electron events are caused by transport-related effects. The spectral fits and the slopes obtained here are unique since it is the first such result obtained close to the source, that is, the IP shock. Particularly, the spectral break observed at high energies (i.e. $> 200$ keV) reported for the first time here is therefore likely a direct consequence of acceleration. 


Both fast-Fermi and SDA mechanisms are candidates for electron acceleration. The characteristics of these mechanisms depend on the conditions at the source, the seed distribution function, and the shock properties (e.g., obliquity, criticality, etc.). The solar wind conditions in reality are constantly changing, altering the conditions for acceleration, favoring a rapid mechanism. Furthermore, it is possible that electrons accelerated in such hot-spots can drift back onto the shock due to the $\textbf{E} \times \textbf{B}$ drift at large distances and undergo further acceleration via a first-order Fermi process.

This letter presents novel observations of relativistic electron beams and demonstrates the efficiency of IP shock acceleration of electrons at 1 AU. The results highlight the importance of shock geometry, large-scale deformations, and the resultant curvature in electron acceleration. This research paves the way for investigating similar phenomena closer to the sun, where shock waves are inherently curved, predominantly near-perpendicular, and are in proximity to dense populations of energetic particles \citep[][]{Jebaraj23b}.

\begin{acknowledgements}
I.C.J and N.D. are grateful for support by the Academy of Finland (SHOCKSEE, grant No.\ 346902). 
We also acknowledge funding by the European Union’s Horizon 2020 research and innovation program under grant agreement No.\ 101004159 (SERPENTINE).
I.C.J acknowledges support from ISSI as part of the visiting scientist program.
I.C.J, V.K, and O.V.A. acknowledge support from the International Space Science Institute (ISSI) in Bern, through ISSI International Team project number 23-575, “Collisionless Shock as a Self-Regulatory System”. 
V.K., acknowledges financial support from CNES through grants, ``Parker Solar Probe'', and ``Solar Orbiter''. 
V.K., and O.V.A. were supported by NASA grants 80NSSC20K0697 and 80NSSC21K1770. O.V.A was partially supported by NSF grant number 1914670, NASA’s Living with a Star (LWS) program (contract 80NSSC20K0218), and NASA grants contracts 80NNSC19K0848, 80NSSC22K0433, 80NSSC22K0522.
A.L. is supported by STFC Consolidated Grant ST/T00018X/1.
A.K. acknowledges financial support from NASA NNN06AA01C (SO-SIS Phase-E, PSP EPI-Lo) contract.
N.W.\ acknowledges support from the Research Foundation - Flanders (FWO-Vlaanderen, fellowship no.\ 1184319N).
D.L. acknowledges support from NASA Living With a Star (LWS) program NNH19ZDA001N-LWS, and the Goddard Space Flight Center Heliophysics Innovation Fund (HIF) program.
I.C.J is grateful for the fruitful discussion with Dr. Jens Pomoell regarding IP shocks. 
Solar Orbiter is a space mission of international collaboration between ESA and NASA, operated by ESA.
The open-source Python packages SEPpy \citep{Palmroos22}, solo-epd-loader \citep{epdloader2023}, and sunpy \citep{Barnes23} have been used in obtaining and processing Solar Orbiter data.
\end{acknowledgements}

\bibliographystyle{aa.bst} 
\bibliography{ref} 

\appendix

\section{Experimental details of the energetic electron measurements}\label{app:all_telescopes}

 \begin{figure}[ht]
 \centering
    \includegraphics[width=0.499\textwidth]{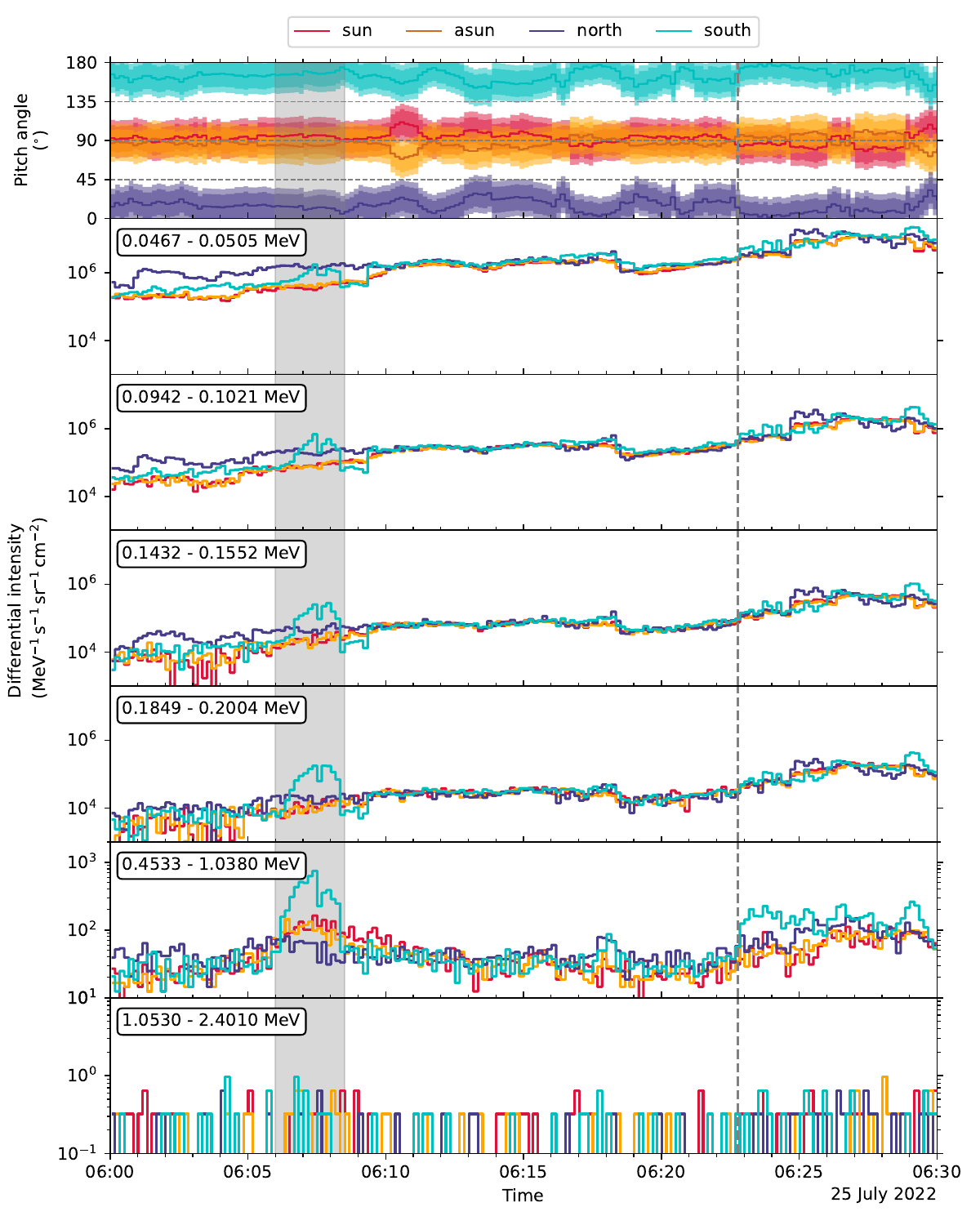}
  \caption{Differential intensities of energetic electrons measured by all four viewing directions of EPT and HET at the time of shock arrival (gray vertical dashed line). EPT intensities were corrected for ion contamination. To be used complementary to Fig.~\ref{Fig:overview}. }
  \label{Fig_app:full_electrons}
  \end{figure}

 \begin{figure}[ht]
 \centering
    \includegraphics[width=0.499\textwidth]{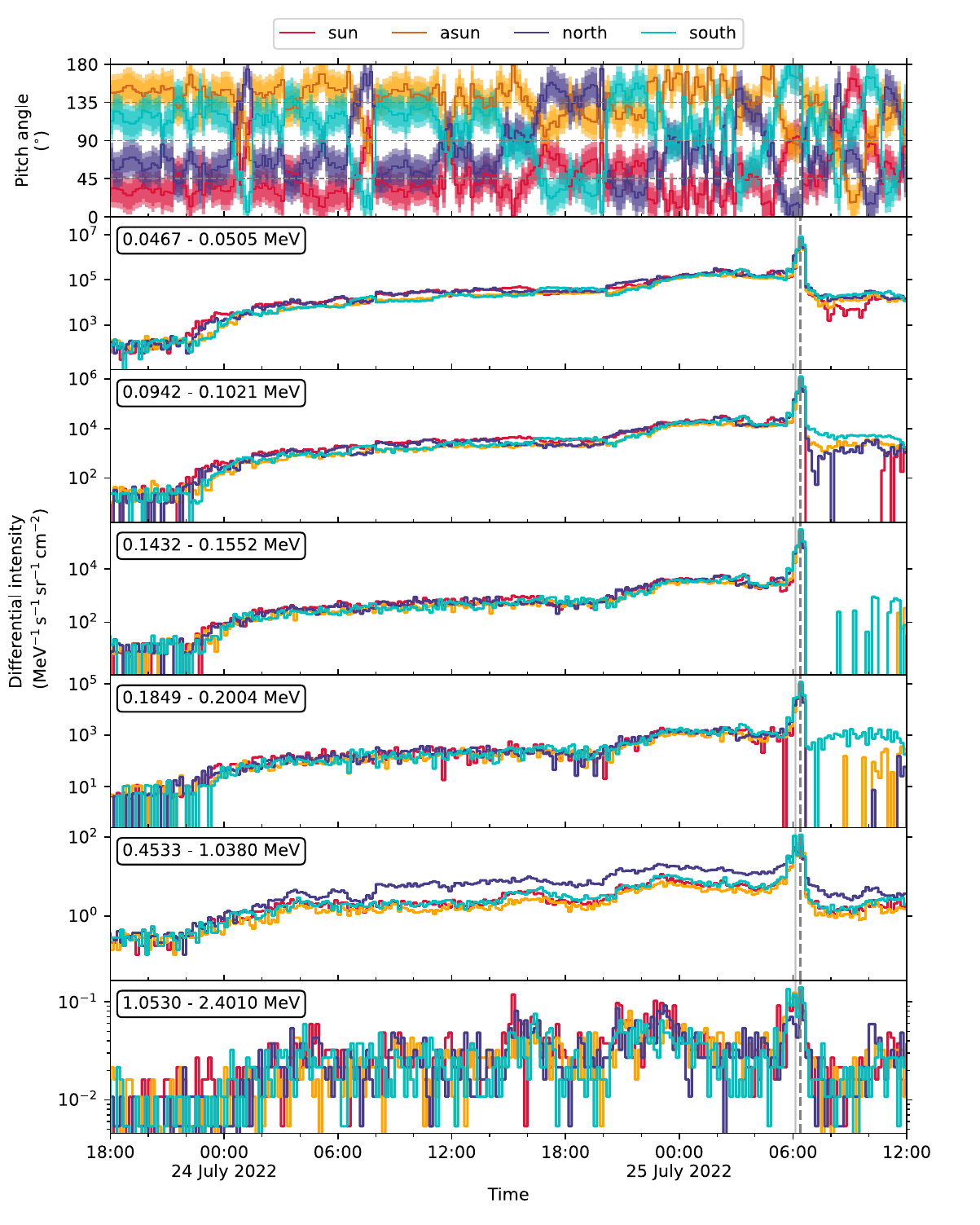}
  \caption{Differential intensities of energetic electrons measured by all four viewing directions of EPT and HET for the whole event time (shock arrival indicated by gray vertical dashed line). EPT intensities were corrected for ion contamination. To be used complementary to Fig.~\ref{Fig:overview}. }
  \label{Fig_app:full_electrons_2}
  \end{figure}

During shock arrival, a steady increase in the differential intensities of the energetic electrons was observed by all four viewing directions of EPT and HET, namely, sun, anti-sun, north, and south, respectively. This letter focuses on the upstream beam observed uniquely by the south telescope. This section provides supportive information for the electrons observed during the event. Complementary to Fig. \ref{Fig:overview} presented in the main text, Fig. \ref{Fig_app:full_electrons} shows the pitch angle coverage and the same energy channels as observed by all four telescopes. The energy channels are the same as the ones presented in the bottom panel of Fig. \ref{Fig:overview}. 

In Fig. \ref{Fig_app:full_electrons} a number of features related to electron anisotropy at the vicinity of the shock aside from the FABs can also be identified. Firstly, an increase in the intensity of relativistic electrons along PA $180^{\circ}$ (second panel from bottom) after the arrival of the shock at 06:22:50 UT. Another one is the bi-drectional beams observed at $\sim$06:18 UT. Identification of such features is possible thanks to the enhanced resolution of the EPD instrumentation.

Figure \ref{Fig_app:full_electrons_2} presents the full electron event from the solar eruption till the shock arrival at PSP. The same energy channels as in Fig. \ref{Fig:overview} and Fig. \ref{Fig_app:full_electrons} are shown together with the pitch angles in the top panel for the three day period between July 23 -- 26, 2022. As with Fig. \ref{Fig_app:full_electrons}, a number of observations can be made about the presence of an energetic storm particle event for electrons which are not the focus of the study. Firstly, the ESP event is seen at all energy ranges presented here which range from near-relativistic to relativistic. The second panel from bottom shows minor increases in the $\sim$0.5 -- 1 MeV electron intensities above the background level. Such relativistic electron observations are extremely rare in IP shocks \citep{Nasrin23}. Secondly, the relativistic electrons (panel above the bottom panel) show a noticeable anisotropy for almost the entire event. 

\subsection{Ion contamination of Solar Orbiter EPD/EPT measurements}\label{app:conta_corr}

\begin{figure*}
    \centering
    \includegraphics[width=0.99\textwidth]{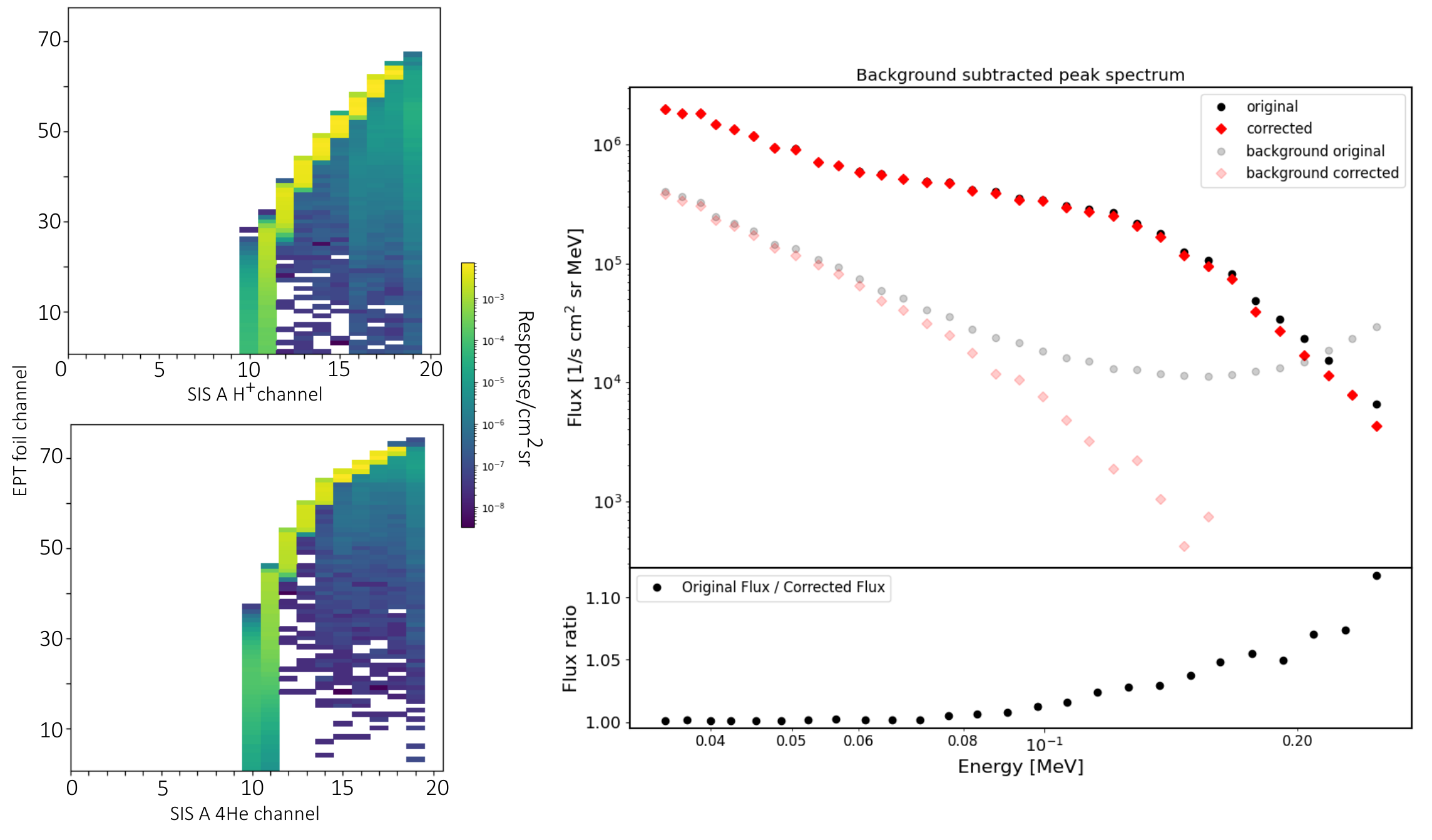}
    \caption{Ion contamination correction for EPD/EPT channels. Response matrices showing the mean response of all EPT foil channels to H (\textit{left top panel}) and $^4$He (\textit{left bottom panel}) in the corresponding energy channels provided by SIS. (\textit{right top panel}) Corrected vs original data of both the electron flux and the background. (\textit{right bottom panel}) Ratio of original vs corrected flux.  }
    \label{fig:EPTxSIS_Response}
\end{figure*}

EPT uses the so called magnet-foil technique to separate and measure electrons and ions. This measuring principle has the inherent characteristic that energetic ions will contaminate the electron channels. To determine reliable electron fluxes, the ion contamination in the electron channels must be taken into account. 
For this work, we use the ion response functions from EPT together with H and $^4$He flux measurements from SIS-A in order to estimate and correct the ion contamination. In particular, we calculate the count rate of contaminating ions $C_{\mathrm{ion}}$ observed in one of the electron channels of EPT by 
\begin{equation}
    C_{\mathrm{ion}} = \sum_x R^{\mathrm{H}}_x I^{\mathrm{H}}_x \Delta E_x + \sum_y R^{\mathrm{He}}_y I^{\mathrm{He}}_y \Delta E_y ,
    \label{Eq:ionResponse2}
\end{equation}
where  $I^{\mathrm{H}}_x$ and $I^{\mathrm{He}}_y$ are the H and $^4$He fluxes observed by SIS in the energy channels x and y, $\Delta E_x$ and $\Delta E_y$ are the energy widths of the channels x and y and $R^{\mathrm{H}}_x$ and $R^{\mathrm{He}}_y$ are the mean responses of EPT for H and $^4$He in the energy ranges of the channels x and y. Figure \ref{fig:EPTxSIS_Response} left panels shows the mean response factors $R^{\mathrm{H}}_x$ and $R^{\mathrm{He}}_y$ for each EPT foil channel in a response matrix.

The ion count rate $C_{\mathrm{ion}}$ in each electron channel can be subtracted from the observed count rate in order to obtain an almost clean electron count rate. This electron count rate can be calibrated into electron fluxes using the electron calibration factors from EPT.

Although this method should in principle produce clean electron fluxes, there are some limitations that need to be considered. SIS and EPT have different fields of view, which can lead to over- or under-correction of EPT electron measurements in the case of anisotropic ion fluxes. Furthermore, the energy coverage and energy resolution of SIS are limited, which also limits the accuracy of the correction. Lastly, all ions other than H and $^{4}$He are neglected here.

For the time periods investigated here, the anisotropy of the ions was so small, that the assumption of isotropic ion fluxes seemed reasonable for the correction of the ion contamination. Ion species other than H and $^4$He were not considered here because their flux is generally small compared to H and $^4$He fluxes and no significant contribution to the observed count rates in EPT is expected.

The correction made based on this method during the investigated period is shown in the right panels of Fig. \ref{fig:EPTxSIS_Response}. The correction is minimal ($\sim$ 1\% or lower) up to 100 keV. This ratio increases to $>$5\% for the energy channels covering $>$ 200 keV energies.

\subsection{Ion contamination of Solar Orbiter EPD/HET measurements} \label{app:het}

   \begin{figure*}
 \centering
    \includegraphics[width=0.85\textwidth]{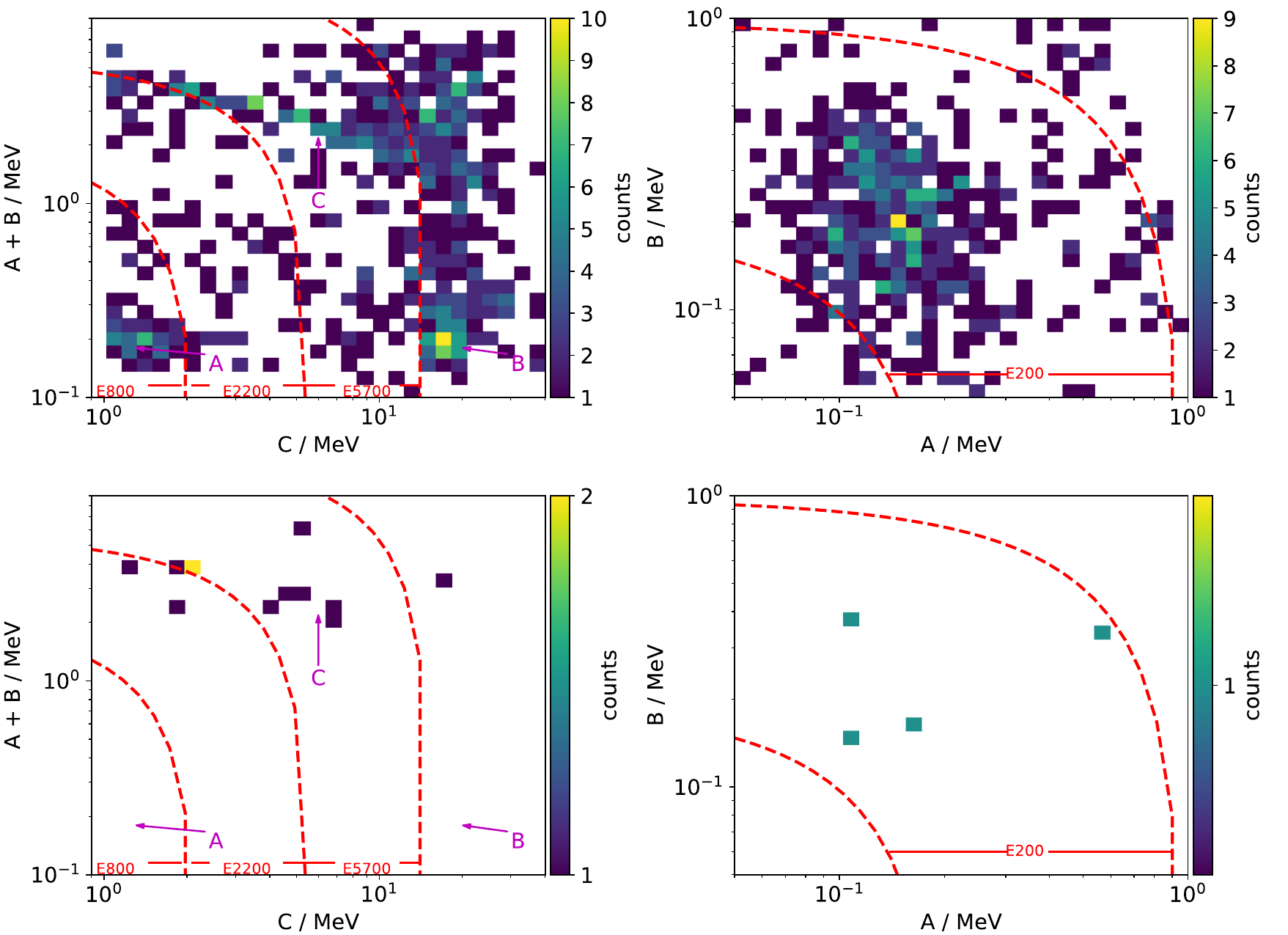}
  \caption{Pulse height analysis (PHA) of HET electrons. (\textit{top left}) PHA data of electrons stopping in detector C, 25~July 2022, 6:00-14:00. (\textit{bottom left}) PHA data of electrons stopping in detector C, 25~July 2022, 6:00-6:10. (\textit{top right}) PHA data of electrons stopping in detector B, 25~July 2022, 6:00-14:00. (\textit{bottom right}) PHA data of electrons stopping in detector B, 25~July 2022, 6:00-6:10. }
  \label{Fig_app:het}
  \end{figure*}
  




The higher energy electron channels of the HET instrument have been found to suffer from proton contamination during strong shock passages. While this issue will be solved in the near future by a software upload from the HET team, careful investigation of the already taken Pulse Height Analysis (PHA) data allows to identify time intervals featuring this contamination. \newline
Figure~\ref{Fig_app:het} top left panel shows PHA data of particles identified by the instrument as electrons penetrating the solid state detectors A and B of HET while stopping in the scintillator C during a prolonged time interval on July, 25th, 2022. This representation (energy loss in A+B vs C) allows to identify several different populations, annotated by A) real electrons which would extend further to the right for higher energies, B) minimal ionizing particles, that is, protons with energies in the GeV range, and C) a track of lower energetic protons. All populations have been reproduced by Monte Carlo simulations. The dashed red lines indicate the energy ranges of the HET electron channel, that is, particles between the right and center red line will be counted in the E5700 channel (the naming scheme refers to \citet{Fleth23}). While the data product has been crafted such that population B) does not spoil the channel, a strong proton contamination due to track C) can be seen during this day. Figure~\ref{Fig_app:het} bottom left panel shows the same representation, however limited to 6:00-6:10, that is, the time of interest for this study. While the PHA statistic is limited here, it is obvious that the higher electron channels are also heavily contaminated by protons during this time series.\newline
Figures \ref{Fig_app:het} top right and bottom right panels show PHA data of particles identified by the instrument as electrons penetrating detector A but stopping in detector B (without reaching the scintillator C), corresponding to HET's lowest electron energy channel employed in this study. In contrast to the ABC coincidence discussed above, no clear track as expected from a possible proton contamination is visible in either of the time series. Furthermore, Monte Carlo simulations do not indicate any possibility of proton contamination in this channel.

\subsection{Velocity dispersion analysis} \label{app:vda}


 \begin{figure}
 \centering
    \includegraphics[width=0.49\textwidth]{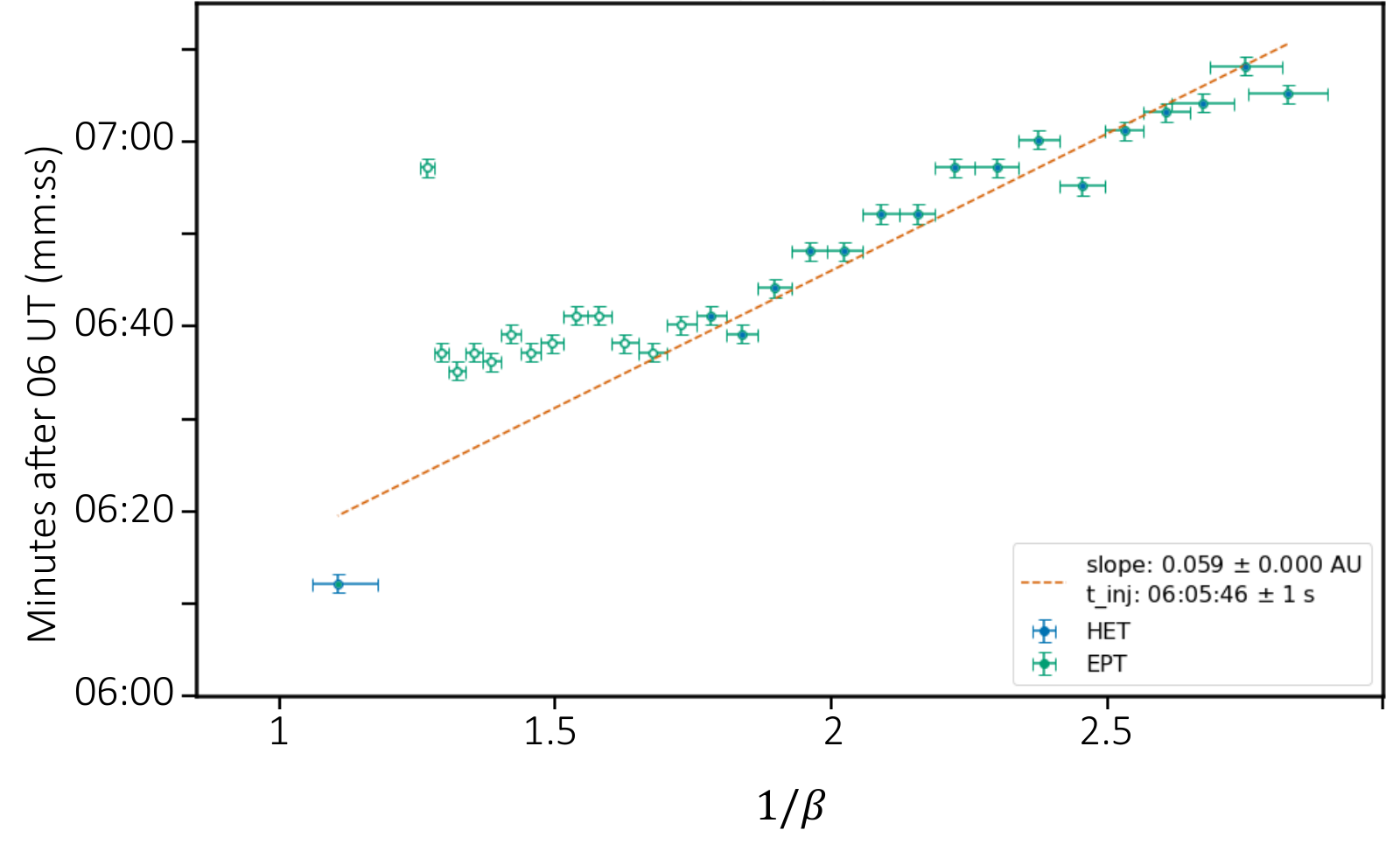}
  \caption{VDA based on onset times of Solar Orbiter EPT and HET electron channels. The horizontal axis shows the inverse of the average unit-less speed of electrons as observed in each channel. The vertical axis presents the determined onset time for each energy channel. Onset times observed by EPT (HET) are marked in green (blue). Horizontal error bars represent the width of the energy channels, and vertical error bars represent the time resolution used to determine the onsets (1 second). The orange line represents the linear fit to the first 16 energy channels of EPT and the first channel of HET, respectively. The onset times that are not considered to the fit have a white middle on the data points.}
  \label{Fig:VDA_temp}
  \end{figure}

Based on the assumption that electrons of all energies were injected at the same time \citep[][]{Vainio13}, the result of the velocity dispersion analysis (VDA) presents an estimation of the injection time of the energetic electrons and the associated path length from the source to the spacecraft. Figure~\ref{Fig:VDA_temp} shows a well-defined velocity dispersion of the beam in the EPT channels $<300$~keV where ion-contamination is low. A good agreement between the EPT channels and the first HET channel is also observed. In this case, the estimated path length is $\sim 13~R_{\odot}$, which is 6\% of the Sun-Earth distance, making the beam a local phenomenon.

It is worth stating that the VDA could only be performed for the onset times of the first peaks, as it was not possible to obtain reliable starting times for the second peaks that occur on top of the decay of the first ones. Also, a VDA performed upon the first peaks themselves exhibited no velocity dispersion. In this regard, the estimation is approximated to the entirety of the FAB. The onset times were identified using the \textit{SEPpy} software package \citep[][]{Palmroos22} which employs the statistical Poisson-CUSUM method \citep[see e.g.,][]{Huttunen2005} to automatically identify the time of the onset.



\section{The relativistic fast-Fermi model for electron acceleration} \label{app:fast_fermi}

The dynamics of particle acceleration in the fast-Fermi model are most effectively described in the de Hoffmann-Teller (HT) frame. In this frame, the upstream plasma bulk velocity is parallel to the magnetic field, leading to no upstream motional (\(\mathbf{-V} \times \mathbf{B}\)) electric field. The velocity transformation to the HT frame from the normal incidence frame (NIF), a shock rest frame where the upstream plasma flow is directed along the shock normal, is given by:

\begin{equation} \label{eq_1}
    \mathbf{u}_{HT} = \frac{\hat{\mathbf{n}} \times (\mathbf{V} \times \mathbf{B})}{\mathbf{B} \cdot \hat{\mathbf{n}}}
\end{equation}

where \( \hat{\mathbf{n}} \) is the shock normal, and \( \mathbf{V} \) and \( \mathbf{B} \) denote the upstream bulk velocity and the magnetic field, respectively. \( \mathbf{u}_{HT} \) lies in the shock plane, parallel to the projection of the upstream magnetic field into this plane. The energization of electrons in the upstream solar wind frame arises from transformations to the HT frame, reflection, and then transformation back to the solar wind frame. Denoting the upstream plasma speed in the HT frame by \( V_{HT} \), it was first demonstrated by \cite{Sonnerup69} that a larger \( V_{HT} \) leads to greater energy gain. The relationship between \( V_{HT} \) and \( V \) can be expressed as:

\begin{equation} \label{eq_2}
    V_{HT} = \frac{V}{\cos\theta_{Bn}}
\end{equation}

An electron entering with a parallel velocity \( v_{\parallel} \)in the upstream plasma frame  will reflect upstream with a parallel velocity of \( - v_{\parallel} + 2V_{HT} \). In relativistic scenarios, this can be described in terms of momentum \( p \) as:

\begin{equation} \label{eq_3}
    \Delta p_{\parallel,HT} = -2 \gamma_{HT} m v_{\parallel, HT} = -2p_{\parallel,HT},
\end{equation}
where, \( \gamma \) denotes the Lorentz factor.
The associated energy change in the upstream plasma frame is:

\begin{equation} \label{eq_4}
    \Delta E = -\Gamma V_{HT} \Delta p_{\parallel,HT} = 2  \Gamma V_{HT} p_{\parallel,HT}
\end{equation}

\begin{figure}
 \centering
    \includegraphics[width=0.49\textwidth]{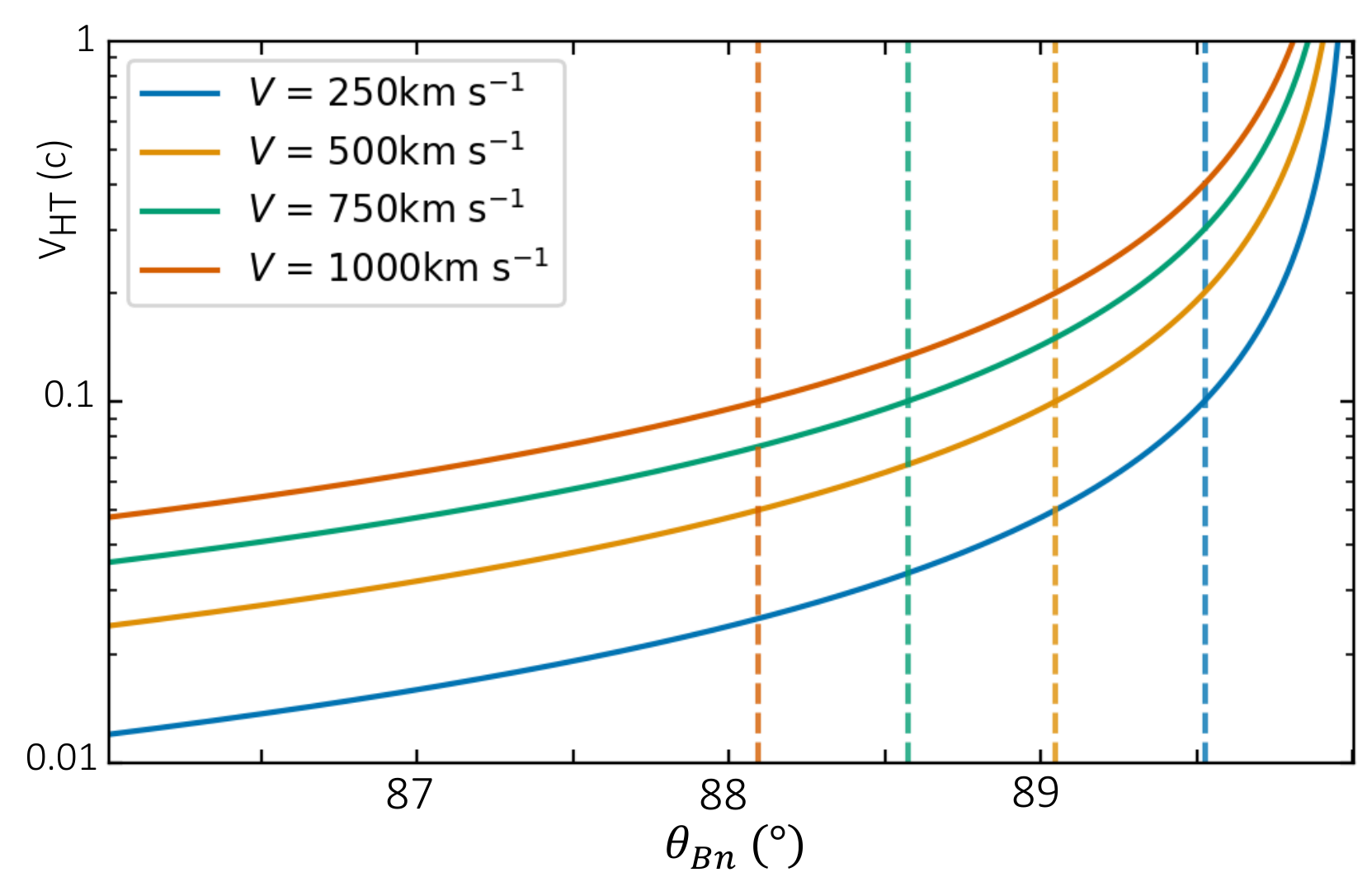}
  \caption{de Hoffmann-Teller transformation velocity (in units of light speed, $c$) for different upstream speeds, $V$ (NIF) as a function of $\theta_{Bn}$. The relationship between these parameters is shown in Eq. \ref{eq_2}. The vertical dashed lines indicate the $\theta_{Bn}$ beyond which the $V_{HT}$ is relativistic. The colors correspond to the different upstream speeds, $V$.}
  \label{Fig_app:probability}
  \end{figure}

Here, \( \Gamma \) represents the Lorentz transformation of \( V_{HT} \), defined as \( \left( 1 - V_{HT}^2/c^2\right)^{-1/2} \). As indicated in Figure \ref{Fig_app:probability}, for \( \theta_{Bn} \) approaching \( 90^{\circ} \), the shock becomes superluminal, making \( \Gamma \) significant. The transformation becomes invalid at \( \theta_{Bn} = 90^{\circ} \).

The change in particle momentum due to reflection is given by Eq. \ref{eq_3}. To comment on the transformation between the HT and the upstream flow frame, one can use:

\begin{eqnarray} \label{eq_5}
    \gamma = \Gamma \left( \gamma_{HT} + p_{\parallel, HT} V_{HT}\right) \\
    p_{\parallel} = \Gamma \left( p_{\parallel, HT} + V_{HT} \gamma_{HT}\right).
\end{eqnarray}

The process of reflection itself happens in the HT frame where the conservation of both the energy and the magnetic moment apply, 

\begin{equation} \label{eq_6}
    p^2_{\parallel , HT} + p^2_{\perp , HT} - e\Phi_{HT} = \text{constant}
\end{equation}

Here, $p$ is the momentum and is divided into $p_{\parallel}$ and $p_{\perp}$ corresponding to parallel and perpendicular momentum. $e\Phi_{HT}$ is the electrostatic potential in the HT and is well known to impact the reflection process \citep[][]{Gedalin04}. 

For relativistic particles, we must consider the following form for energy:

\begin{equation} \label{eq:B_4}
    E = \sqrt{m^2c^4 + p_{HT}^2c^2} + e\Phi_{HT}
\end{equation}

Next, the magnetic moment of the electron may be given as,

\begin{equation} \label{eq:B_5}
    I_B = \frac{p^2_{\perp , HT}}{B}
\end{equation}

This implies that the particle continues to gain perpendicular momentum till a characteristic location at which $p_{\parallel ,HT} = 0$. This is chosen in such a way that it is implicitly assumed that $e\Phi_{HT}$ and $B$ are maximum in the same region, i.e. the reflection point. An important point to consider here is the value of $e\Phi_{HT}$ in comparison with the $p_{\parallel}$ of the electron. A smaller $e\Phi_{HT}$ would be insignificant to the incident electron population. This would immediately ensure that the core electrons in any distribution will not be reflected. 

It is now possible to make use of the Eq.~\ref{eq:B_5} and the adiabatic invariance of a relativistic electron in the upstream plasma ($-\infty$) to that of one at the ramp under the influence of the magnetic mirror ($B_{overshoot}$), 

\begin{eqnarray} \label{eq:B_10}
    \frac{p^2_{\perp ,HT} (- \infty)}{B} = \frac{p^2_{\perp ,HT} (overshoot)}{B_{overshoot}} \nonumber \\
    p^2_{\perp ,HT} (overshoot) = p^2_{\perp ,HT} (-\infty) \frac{B_{overshoot}}{B}
\end{eqnarray}


Substituting Eq.~\ref{eq:B_4} in Eq.~\ref{eq:B_10} one gets,

\begin{eqnarray} \label{eq:B_11}
    \sqrt{m^2c^4 + p^2_{\parallel ,HT} (-\infty)c^2 + c^2 p^2_{\perp ,HT} (-\infty)} = \nonumber \\
    \sqrt{m^2c^4 + c^2 p^2_{\perp ,HT} (-\infty) \frac{B_{overshoot}}{B}} + e\Delta\Phi_{HT}
\end{eqnarray}

If we were to work in the frame where $\textbf{V} || \hat{\textbf{n}}$, i.e. the NIF, then there are two electric field components. One associated with the motion of the plasma $E_y$, and another associated with the shock transition, $E_x$. Transforming into the HT frame then gives,

\begin{equation}
    E_{x , HT} = E_x + \left(\frac{\textbf{V}_{HT}}{c} \times \textbf{B}\right)
\end{equation}

Using this, \cite{Goodrich84} showed that, 

\begin{eqnarray}
    eE_{x , HT} = \nonumber \\
    - \frac{\partial}{\partial x} \left[ \frac{m}{2} \left(v^{2}_y + v^{2}_z\right) + m V_{HT} v_z\right] - \frac{1}{n} \frac{\partial p_e}{\partial x} = \nonumber \\
     -e \frac{\partial \Phi_{HT}}{\partial x}
\end{eqnarray}

$v_x$ and $v_y$ are small compared to the upstream electron thermal speed and so the potential jump $\Delta \Phi_{HT}$ is equal to the change in electron temperature, 

\begin{equation}
    e\Delta \Phi_{HT} = \Delta T_e
\end{equation}

Here, $T_e$ is in units of eV. The change in electron temperature across the shock is usually in the order of 10s of eV to about 100 eV \citep[][]{Lembege04}. Several more proxies can also be used to constrain the electrostatic potential as done so in \cite{Hanson19} based on the methods established in \cite{Gedalin04}. They are based on ion velocity, ion density, pressure balance, and the electron equation of state. In the case of the shock analyzed here, the potential was found to be between 100 -- 200 eV. This estimate is large by more than a factor of 5 compared to the IP shock evaluated by \cite{Hanson19}. The effect of the electrostatic potential is that the acceleration is completely restricted to only the tail of the seed distribution \citep[][]{Leroy84}.







It is quite clear from this entire ordeal that the most important variable in Eq.~\ref{eq:B_11} is the mirror ratio $B_{overshoot}/B$. In the case of a super-critical quasi-perpendicular shock, the overshoot magnetic field can exceed the simple discontinuity limit (MHD) of 4 (for a gas with adiabatic index $5/3$). The reflected electrons therefore form a loss cone distribution which is dependent upon $V_{HT}$ and $e \Delta\Phi_{HT}$. Furthermore, the opening angle, $\alpha$ of the so-called ``loss-cone'' itself is set by the ratio of upstream to downstream magnetic field. The loss-cone angle is such that the resultant distribution is purely made of particles which exceed a certain pitch angle and perpendicular velocity. 

Ideally, this may never exceed a certain limit under MHD conservation laws, however, for the considerations made here where the presence of $B_{overshoot}$ is acknowledged, $\alpha = 1/\sin(B/B_{overshoot})$. As such, it is evident that the upstream distribution of electrons plays an important role in determining whether the acceleration is significant or not. The ``seed'' distribution is based upon the in-situ spacecraft measurements by \cite{Lin74} who showed that the electrons have an enhanced supra-thermal tail in the solar wind as opposed to a purely thermal Maxwellian distribution. The supra-thermal tail can be represented by a single kappa distribution can be defined as,

\begin{equation}
    f = A \cdot \left[1 + \frac{E}{\kappa E_{\kappa}}\right]^{-\kappa-1}
\end{equation}

where $A$ is the normalization constant, and $E$ is the kinetic energy \citep[][]{Maksimovic97} in the seed distribution function. For energies where $E \rightarrow E_{\kappa}$, the distribution behaves as a Maxwellian 


The efficiency of acceleration without relativistic effects was originally estimated by \cite{KraussV1989b}. Figure~\ref{Fig_app:probability} plots Eq.~\ref{eq_2} which shows that as $\theta_{Bn}$ increases above $86^{\circ}$, the HT transformation velocity alone may result in an exponential energy gain of the particle. Therefore, the role of $\theta_{Bn}$ is crucial, as only the electrons which exceed $V_{HT}$ may interact with the shock, resulting in a loss-cone distribution. At a certain limit ($\theta_{Bn} = 90^{\circ}$), none of the electrons are reflected, and they are transmitted downstream \citep{Ball2001}. This aspect of the acceleration mechanism results in the formation of ``dilute'' beams, that is, beams with very small $n_{b}/n$ (ratio of beam density to the background).

\begin{figure}
 \centering
    \includegraphics[width=0.49\textwidth]{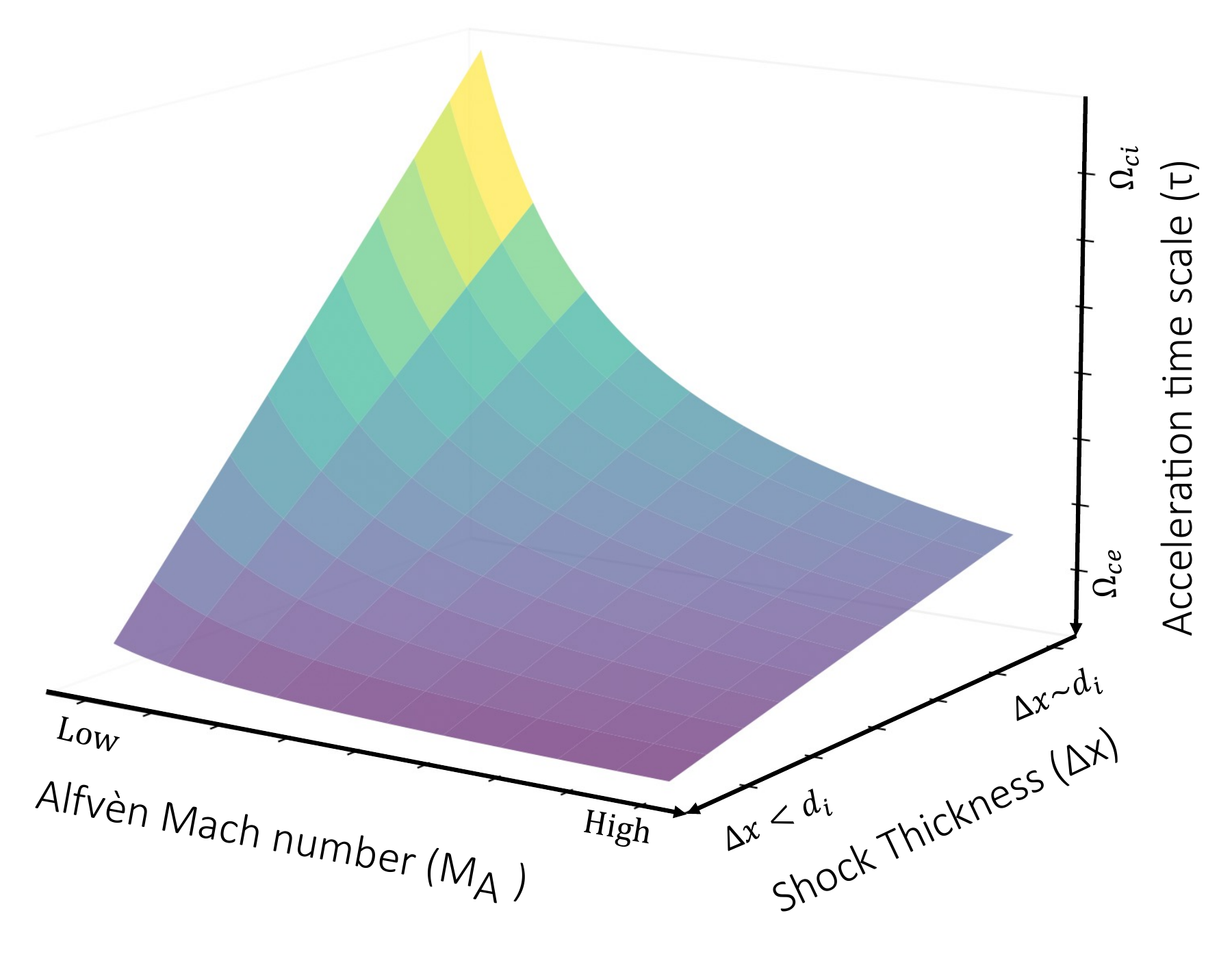}
  \caption{Acceleration time scale ($\tau$) with respect to shock thickness ($\Delta x$) and Alfvèn Mach number ($M_A$). Based on Eq. \ref{eq:ts}. Here, the color coding follows a gradient from light to dark which represents decreasing excursion time (faster acceleration) of the electron.}
  \label{Fig_app:acc_ts}
  \end{figure}

Two important issues not discussed here are the so-called ``injection'' problem and the acceleration time scales. The first issue is fundamental and implies that shock acceleration requires a pre-existing population of supra-thermal particles which can participate in the process. Here, it is overcome by the fact that the acceleration mechanism ignores the thermal electrons and only acts on the enhanced tail population of the kappa distribution. The choice of a kappa distribution further simplifies the issue where the supra-thermal tail is enhanced. The reflection time scale for fast-Fermi is dependent upon several factors. The most important of which are the shock strength $M_A$, the thickness of the shock $\Delta x$, and $B_{overshoot}$. Under the non-relativistic limit for a weak shock with $\Delta x$ in the order of the ion-inertial length ($d_i = c/\omega_{pi}$) and for a quasi-perpendicular shock, this may be approximated to depend directly on the ion gyro period \citep[$\Omega_{ci}$, ][]{Leroy84}, 

\begin{equation} \label{eq:ts}
    \tau \sim 4 \Omega_{ci}^{-1} \frac{\Delta x}{M_A} \frac{B_{overshoot}}{\Delta B} \left[1 + \frac{2}{3} \frac{B_{overshoot}}{B}\right]
\end{equation}

here, $\Delta B = B_{overhsoot} - B$. The relationship is visualized in Fig. \ref{Fig_app:acc_ts} which shows that for a strong shock with a large magnetic overshoot, its thickness is $<d_i$ \citep[][]{Hobara10}. The issue of acceleration time scales becomes important when distinguishing between the various acceleration mechanisms such as fast-Fermi, and first-order Fermi acceleration. The latter due to its stochastic nature, happens over longer time periods. Meanwhile, the former in the relativistic consideration is in the order of the electron gyro period and therefore at least two orders of magnitude faster than in the non-relativistic case. 

A matter of caution in the coordinate transformations considered here is that the reflection process can be treated either purely in the HT frame or the NIF. However, there are frame specific effects such as the transmitted electrons gaining energy in the HT frame due to the potential, while being unaffected in the NIF due to the convective electric field. If acceleration is purely treated in the HT frame, then the description provided in this section is satisfactory, that is, energy is gained via frame transformation. However, if purely treated in the NIF, the motional electric field ensures that the particles enter a gradient drift motion. In such a case, particles will be seen in the direction of the non-coplanar component of the magnetic field.

\end{document}